\documentclass[twocolumn,aps, prd, amsmath, amssymb, bm]{revtex4}
\usepackage{times}
\usepackage{graphicx}
\usepackage{verbatim}
\usepackage{tikz}
\usepackage{color}

\usepackage[T1]{fontenc}
\usepackage{aecompl}
\usepackage{calligra}
\DeclareMathAlphabet{\mathcalligra}{T1}{calligra}{m}{n}
\DeclareFontShape{T1}{calligra}{m}{n}{<->s*[2.2]callig15}{}

\usepackage{hyperref}
\hypersetup{
    pdfnewwindow=true,      
    colorlinks=true,       
    linkcolor=blue,          
    citecolor=blue,        
    filecolor=blue,      
    urlcolor=blue           
}

\def\orb{\rm{orb}}

\def\GW{\rm{GW}}

\def\Msun{{M_{\odot}}}
\def\mbh{\BHcirc}

\def\GW{\rm{GW}}

\def\HoR{\mathcal{H}}
\def\spin{\mathcal{S}}
\def\BHcirc{\tikz\draw[black,fill=black] (0,0) circle (.25ex);}

\begin{document}
\title[A Single Progenitor Model]{A Single Progenitor Model for GW150914 and GW170104}
\author{Daniel J. D'Orazio}
\email{daniel.dorazio@cfa.harvard.edu}
\affiliation{Astronomy Department, Harvard University, 60 Garden Street Cambridge, MA 01238, USA}

\author{Abraham Loeb}
\email{aloeb@cfa.harvard.edu}
\affiliation{Astronomy Department, Harvard University, 60 Garden Street Cambridge, MA 01238, USA}

\begin{abstract} 
The merger of stellar-mass black holes (BHs) is not expected to generate
detectable electromagnetic (EM) emission. However, the gravitational
wave (GW) events GW150914 and GW170104, detected by the Laser Interferometer
Gravitational Wave Observatory (LIGO) to be the result of merging, $\sim 60
\Msun$ binary black holes (BBHs), each have claimed coincident gamma-ray
emission. Motivated by the intriguing possibility of an EM counterpart to BBH
mergers, we construct a model that can reproduce the observed EM and GW
signals for GW150914- and GW170104-like events, from a single-star progenitor.
Following \textcite{Loeb:2016}, we envision a massive, rapidly rotating star within
which a rotating bar instability fractures the core into two overdensities
that fragment into clumps which merge to form BHs in a tight binary
with arbitrary spin-orbit alignment. Once formed, the BBH inspirals due to gas
and gravitational-wave drag until tidal forces trigger strong feeding of the
BHs with the surrounding stellar-density gas about 10 seconds before merger.
The resulting giga-Eddington accretion peak launches a jet that breaks out of
the progenitor star and drives a powerful outflow that clears the gas from the
orbit of the binary within one second, preserving the vacuum GW waveform in
the LIGO band. The single-progenitor scenario predicts the existence of
variability of the gamma-ray burst, modulated at the $\sim 0.2$ second
chirping period of the BBH due to relativistic Doppler boost. The jet breakout
should be accompanied by a low-luminosity supernova. Finally, because the BBHs
of the single progenitor model do not exist at large separations, they will
not be detectable in the low frequency gravitational wave band of the Laser
Interferometer Space Antenna (LISA). Hence, the single-progenitor BBHs will be
unambiguously discernible from BBHs formed through alternate, double-progenitor 
evolution scenarios.
\end{abstract}

\maketitle

\section{Introduction}

The Laser Interferometer Gravitational Wave Observatory (LIGO) has
conclusively detected gravitational waves (GWs) from the merger of two black
holes (BHs) in five different systems \citep{GW150914:2016, GW151226:2016,
GW170104:2017, GW170608, GW170814}. In addition to its notoriety as the first
detected GW signal, GW150914 also made waves for being a peculiarly
\footnote{Though not completely unexpected \citep{Kinugawa+2014,
RodRasio+2015}.} high mass system consisting, before merger, of two nearly
equal mass BHs adding up to $\sim 65 \Msun$ \citep{GW150914:Params:2016}. The
addition of GW170104 and GW170814, similarly high mass, $\sim 50 \Msun$,
binaries with nearly equal mass components, has hinted that such high mass,
near-unity mass ratio systems may be common.

Perhaps more interesting than LIGO's observation of such unexpected
systems is the possibility that two out of the three are associated with an
electromagnetic (EM) counterpart. While no electromagnetic counterpart is
expected from the merger of stellar-mass BHs \footnote{$\lesssim 100 \Msun$ as
opposed to super-massive  $\gtrsim 10^5 \Msun$.} \citep[see][]{Lyutikov:2016},
both GW150914 and GW170104 have been associated with gamma-ray emission
carrying total isotropic energy of $\sim 10^{49} - 10^{50}$ ergs, and
occurring within half of a second from the peak of the gravitational wave
strain \citep{Connaughton+2016, AGILE:Verrecchia+2017}.

We proceed by assuming the gamma-ray transients are indeed connected to the GW
events, and ask what could be their origin? While exotic physics, such as
highly charged BHs \citep[\textit{e.g.},][]{ZhangQQ:2016, Fraschetti:2016}
could be conjectured, we consider more standard astrophysical scenarios. In
all such scenarios, the generation of $\sim10^{49}$ ergs of energy must
correspond to a giga-Eddington event; a $30 \Msun$ BH must accrete at $\sim 3
\times 10^{9}$ times the Eddington rate for one second, or equivalently,
$10^{-4} \Msun$ must be accreted at $10 \%$ efficiency within one second in
order to achieve these energies.

The standard, double-progenitor binary black hole (BBH) formation channels:
(i) isolated evolution of binary systems in the field
\citep[\textit{e.g.},][]{VossTauris:2003, Dominik+2012, Dominik+2013,
Belczynski+2014,ZevinRasio+2017} and (ii) dynamical capture in clusters
\citep[\textit{e.g.},][]{ZevinRasio+2017,Rodriguez+2016, StoneFAU:2017,
Bartos+2017}, do not naturally allow for this much gas to be present at the
time of merger, though recently a number of models have been put forward to
challenge this \citep{Perna+2016, Loeb:2016, Woosley:2016, deMinkKing:2017,
Janiuk+2017}.

Rather than consider possible scenarios for generation of high density gas in
the standard, double-progenitor paradigm, \textcite{Loeb:2016} pointed out that a
single progenitor model (previously studied by \citep{Fryer+2001,
Reisswig+2013}) can naturally provide the gas densities needed to power the
putative gamma-ray transient. In this model, a rotating bar instability forms
in the core of a massive rapidly rotating star, forming a dumbbell
configuration that fissures into the two proto-BHs, which eventually merge in
the LIGO band powering a giga-Eddington accretion burst that results in a
collapsar-type event \citep[\textit{e.g.},][]{MacFadyenWoosley:1999}, possibly
powering a gamma-ray transient.

While providing the correct energies of emission, later work pointed out that:
(i) gas drag on the BHs inside the collapsing star will unmistakably alter the
GW wave form detected by LIGO \citep{Dai+2016, FedrowOtt+2017}, and (ii) due
to the $\sim10$ second jet breakout timescale in the collapsar model, the time
delay between EM and GW signals would be longer than the observed $\sim \pm
0.5$ seconds \citep{Woosley:2016}. In addition to these issues, the model
would naively predict BHs with spins that are aligned with the binary orbital
angular momentum while GW170104 does not show evidence for significant
aligned BH and binary orbital angular momenta \cite[See also
Ref.][]{Kushnir+2016}.

Here we present a single progenitor model for GW150914- and GW170104-like
events in which the above issues are alleviated. We consider a model similar
to that of \textcite{Loeb:2016}, but where tidal forcing of the binary drives a 
giga-Eddington accretion event $\sim10$ seconds before merger, driving a powerful
outflow that: (i) clears the gas surrounding the binary before it reaches the
LIGO band, and (ii) can alter the time delay between EM and GW signatures to
match the observed $\sim \pm 0.5$ second shift from the peak of the LIGO signal.

Also new to the model, we consider a formation scenario for the BHs within the
massive progenitor star that would allow BH spin misalignment (though this is
not required by GW observations). As the rotational bar instability ensues, each
end of the rotating dumbbell can fragment into multiple clumps with Jeans mass
of order a solar mass. As the relaxation time of these clumps is of order a
dynamical time, the clumps would quickly randomize their angular momenta
before merging into a $30 \Msun$ BH, allowing BHs with spins misaligned with
the orbital angular momentum. Additional impacts on the BH after formation can
tilt its spin similarly to the way the spin axis of Uranus is tilted by
asteroid impacts in the early solar system
\citep[\textit{e.g.},][]{Morbidelli+2012}.

While some aspects of the above processes are uncertain, including even the
association of the GWs and gamma-rays themselves, we  stress that the single
progenitor model put forth in this article is a real possibility that carries
with it predictions that would discern it from other double progenitor
scenarios:
	(i) BBHs formed in our scenario will not exist at large enough 
	separations to emit GWs detectable by the Laser Interferometer Space 
	Antenna \citep[LISA;][]{LISA:2017}, as suggested by 
	\citep{Sesana:2016, Seto:2016} for GW150914;
	(ii) Accompanying the merger should be a faint supernova;
	(iii) Because the gamma-ray burst (GRB)-like outflow occurs before merger, 
	the chirping orbital frequency of the binary should be imprinted as 
	variability on the gamma-ray lightcurve;
	(iv) The delay time between GWs and the short gamma-ray transient is dependent
	on binary parameters as well as uncertain hydrodynamics. If future work
	can better pin down the latter, then GW observations that measure binary
	parameters would constrain theoretical models for the EM time delay.

\section{Summary of observations}

We first summarize the gravitational and electromagnetic observations of the
two high-mass BBH LIGO systems with claimed gamma-ray counterparts. Most
relevant to our model are the gamma-ray burst durations, energies, and time
delays with respect to the GW peak, as well as the BH masses, and the
alignment of BH spin relative to the line of sight and to the orbital angular
momentum. Because our goal is to characterize the putative EM counterparts, we
only summarize the relevant claimed EM detections and do not present an
extensive summary of all the EM follow up surveys.

\subsection{GW150914} 

The gravitational wave event GW150914 is due to the merger of two BHs of
masses $36.2^{+5.2}_{-3.8} \Msun$ and $29.1^{+3.7}_{-4.4} \Msun$. The
dimensionless  spin parameter is $\spin_1 = 0.32^{+0.49}_{-0.29}$ for the
primary and $\spin_2 = 0.44^{+0.50}_{-0.40}$ for the secondary. The spin
orientation is not strongly constrained, but if one assumes that the pre-
merger spins are aligned with the binary orbital angular momentum, then
$\spin_1 < 0.2$ and $\spin_2 < 0.3$ with $90 \%$ probability
\citep{GW150914:Params:2016, LIGO_BBHO1:2016}. It is strongly disfavored that
the binary orbital angular momentum is misaligned with the line of sight; the
probability that the angle between the total binary orbital angular momentum
and the line of sight is between $45^{\circ}$ and $135^{\circ}$ degrees is
$0.35$. The peak value of the source-orientation probability distribution
function is $160^{\circ}$, $20^{\circ}$ from  anti- alignment with the line of
sight.

The Gamma-ray Burst Monitor (GBM) on board the Fermi satellite claimed a
($2.9\sigma$) detection of a gamma-ray transient $0.4$ seconds after the
merger time recorded in gravitational waves and consistent with a weak
short gamma-ray burst. The transient lasted $1$ second, and at the
gravitational wave inferred luminosity distance of $410$ Mpc, a total energy
of $1.8^{+1.5}_{-1.0} \times 10^{49}$ ergs was radiated between 1 keV and 10
MeV \citep{Connaughton+2016, Connaughton+2018arXiv_Fermi2}. The
INTEGRAL/SPI-ACS instrument does not detect a coincident gamma-ray signal in
the harder, 75 keV-100 MeV range \citep{INTEGRAL:nullGW150914}.

\subsection{GW170104}
The gravitational wave event GW170104 is due to the merger of two BHs
of masses $31.2^{+8.4}_{-6.0} \Msun$ and $19.4^{+5.3}_{-5.9} \Msun$. The
dimensionless spin parameters of the individual BHs before merger are not
strongly constrained, but large values that are aligned with the binary
angular orbital momentum are disfavored \citep{GW170104:2017}. The binary
orbital angular momentum inclination to the line of sight is not well
constrained with broad probability peaks at face-on and edge-on inclinations.

A gamma-ray transient was detected at the $\sim2.5 \sigma$ level, $0.46 \pm
0.05$ seconds before the GW170104 merger event and lasting $32$ms. The
luminosity and fluence of this event was also consistent with a weak short
gamma-ray burst. At the gravitational wave inferred luminosity distance of
$880$ Mpc, the total energy in the $0.4 - 40$ MeV band is $E_{\rm{iso}} \sim
8.3 \times 10^{48}$ erg corresponding to an isotropic luminosity of
$L_{\rm{iso}} \sim 2.6 \times 10^{50}$ erg s$^{-1}$
\citep{AGILE:Verrecchia+2017}.\footnote{The uncertainty in the
luminosity distances for both events discussed here is large:
$410^{+160}_{-180}$ Mpc (updated in \textcite{LIGO_BBHO1:2016} to
$420^{+150}_{-180}$ Mpc) and $880^{+450}_{-390}$ Mpc for GW150914 and GW170104
respectively.}

Neither the Fermi GBM (10 KeV - 1MeV), the Fermi Large Area Telescope (0.1-1
GeV), the AstroSat-CZTI ($>100$ KeV), or INTEGRAL SPI-ACS reported a detection
of a transient similar to the AGILE detection \citep{Fermi:GW170104:2017,
Bhalerao+2017, INTEGRAL_GW170104:2017}. ATLAS and Pan-STARRS did, however,
report the detection of a GRB afterglow candidate ATLAS17aeu in the GW170104
error circle $23$ hours after the GW event, but we do not consider any
connection to GW170104 here since its inferred host galaxy is likely at a
redshift larger than the GW source \citep{Stalder+2017, Bhalerao+2017}.

In summary, both of the above events consisted of nearly equal mass BBHs of
order $(30+30) \Msun$ whose merger might have coincided with a gamma-ray
transient with total isotropic energy of order $10^{49}$ ergs. While the BH
spin alignments are poorly constrained, the BH spins are consistent with being
aligned towards the observer's line of sight, so the possibly beamed signal
described below could be pointed toward the observer. While alignment of the
BH spins with the binary orbital angular momentum is not ruled out, it is
disfavored for large values of the spin magnitude.

A third, high-mass, near-unity mass ratio BBH detected by LIGO,
GW170814, has no claimed EM counterpart \citep{GW170814}. As the binary and
spin orientations of the LIGO BBH events are poorly constrained, we cannot say
whether or not this can be explained by the viewing angle.

\section{Single Progenitor Model}

\subsection{BBH formation and spin-orbit alignment}

We consider a single, massive $\gtrsim 250 \Msun$, rapidly rotating, 
low-metallicity star as the progenitor of GW150914- and GW170104-like BBH systems.
Such a star would be the natural outcome of the merger of a massive, tight
binary system with a common envelope \citep{deMink+2014, Hwang+2015,
MandelMink:2016}.

Furthermore, such massive stars are expected to form in nearly equal mass
ratio tight binaries and merge within a Hubble time at a rate comparable to
the low end of the BBH merger rate inferred by LIGO, $\sim 10$ Gpc$^{-3}$
yr$^{-1}$ \citep{GW170104:2017}. Ref. \citep{MandelMink:2016} uses a Kroupa
initial mass function (IMF) to estimate the merger rate of $\gtrsim60 \Msun$
stars to be $\sim 20$ Gpc$^{-3}$ yr$^{-1}$. If we simply extend the back-of-
the-envelope argument made by Ref. \citep{MandelMink:2016} to only consider
stars above $125 \Msun$ (assuming that they exist), and assume that such
binaries form in nearly equal mass ratio pairs (see \citep{Hwang+2015}), then
because the Kroupa IMF scales as a $-2.35$ power law in mass, the decrease in
the inferred merger rate drops by only a factor of $(60/125)^{-1.35} \sim 3$.
Considering further that only three of the five LIGO detections are of the
proposed single progenitor type put forth in our model, the rate of
stellar mergers above $\sim 125 \Msun$ is not inconsistent with the rate of
very massive, nearly equal mass ratio BBH mergers inferred by LIGO.

We require the total stellar mass to be above $\sim250 \Msun$ so that stellar
collapse is not subject to the pair instability supernova mechanism, causing
the star to explode, leaving behind no progenitor, or pulsating and losing too
much mass to be the progenitor of a $\sim 60 \Msun$ BBH
\citep[\textit{e.g.},][]{Woosley:2016, Woosley:2017:PPiSN, WooHeger:2015}.

The angular momentum of the star must be below the break-up value of the star,
but also above that of the centrifugal barrier which sets the initial
separation of the BBHs. As in Ref. \citep{Loeb:2016}, we require that the
initial separation $a_0$ of the BBH be large enough to not disturb the LIGO
observations ($\sim 10M$). Additionally, in this model, we require that $a_0$
also be greater than the binary separation at which our EM mechanism turns on,
which we describe below occurs around $20 r_G$ (where $r_G \equiv GM/c^2$
for M the total BBH mass). Conservatively, we require $a_0 \geq 50
r_G$ to constrain the angular momentum budget of the star,
\begin{equation}
1 > \frac{ \Omega R^2_* }{ j_{\rm{max}} }  \gtrsim 0.01 \left( \frac{R_*}{R_c} \right)^2 \left( \frac{M_*}{300 \Msun} \right)^{-3/4},
\label{Eq:jojmx} 
\end{equation}
where, as in Ref. \citep{Loeb:2016}, we posit a star with constant angular
velocity $\Omega$ and angular momentum profile $j = \Omega r^2$, $R_c$ is the
radius of the core that collapses to create the BBH with initial separation
$a_0 \ll R_c$,  $j_{\rm{max}}$ is the angular momentum corresponding to break
up, $M_*$ is the stellar progenitor total mass, and $R_*$ is the progenitor
radius. For a more massive star and a larger required centrifugal barrier than
in the model of Ref. \citep{Loeb:2016}, we arrive at the same result as Eq.
(5) of Ref. \citep{Loeb:2016}.

We note, however, that 1D simulations by \textcite{Heger+2005} and
\textcite{Woosley:2016} find that braking of stellar rotation via magnetic
torques and mass loss could slow the rotation of such massive stars below the
required minimum value to create the BBH. The final fate of the stellar core's
angular momentum, however, is sensitive to the uncertain mass loss rates and
magnetic field implementations used in these 1D calculations. We note this
potential complication but proceed by considering the case where the star can
collapse with the required angular momentum.

The core of the rapidly rotating, collapsing star will become unstable to a
rotating bar instability \citep{Fryer+2001, Reisswig+2013}. The bar will form
into a dumbbell configuration, within which the two BHs will form at either
end. We envision a formation scenario where the gravitationally unstable gas
in each end of the dumbbell fragments in to multiple clumps of mass and size
given approximately by the Jeans criterion,
\begin{eqnarray}
\frac{M_J}{30 \Msun} &\approx& 0.036 \left( \frac{T}{10^9 \rm{K} }\right)^{3/2} \left( \frac{\rho}{10^8 \rm{cm}^{-3} }\right)^{-1/2}  \\
\frac{\Lambda_J c^2}{30 G \Msun} &\approx& 25 \left( \frac{T}{10^9 \rm{K} }\right)^{1/2} \left( \frac{\rho}{10^8 \rm{cm}^{-3} }\right)^{-1/2} ,
\label{Eq:Jean}
\end{eqnarray}
where a typical core density and temperature is estimated from the models of
\textcite{Fryer+2001}. \footnote{If the temperature in the fragmentation region
is closer to $10^{10}$K, at the same density of $\sim10^8$ g cm$^{-3}$, then
the Jeans mass is closer to $30\Msun$ and we do not expect a swarm of clumps
to form, rather the BH will be formed with spin angular momentum aligned with
the binary orbital angular momentum.}  The presence of gravitational
perturbations and shearing forces in the rotating collapsing clumps will alter
the instability criterion away from the simple Jeans approximation. However,
considering even the uncertainty of the temperature and density in each
collapsing clump, a more complex treatment of fluid and gravitational
instability in the collapsing star is beyond the scope of this study.

Each swarm of tens of $\sim \Msun$ clumps (one swarm at either end of the
dumbbell configuration) will be born with the same orbital angular momentum
and spin angular momentum, but will interact with itself gravitationally and
be slowed via gas drag. For a swarm of $N \sim 30\Msun / M_J \sim 30$ clumps,
the relaxation time of the proto-BH swarm is \citep[\textit{e.g.}, \S
1.21 of Ref.][]{B&T2008},
\begin{equation}
t_{\rm{relax}} = \frac{N}{ 8 \rm{ln}{N} }\Omega^{-1}_{\rm{swarm}} \sim \Omega^{-1}_{\rm{swarm}} = \sqrt{\frac{\left(R_c/2\right)^{3} }{G M_{\mbh}}},
\end{equation} 
equal roughly to the dynamical time of the swarm,
$\Omega^{-1}_{\rm{swarm}}$. Here $M_{\mbh}$ is the mass of the single
BH formed by the swarm and we assume a maximum extent of each swarm to be half
the core radius. Assuming that the clumps have a size smaller than the Jeans
length, we can compare the relaxation time to the time until the first
collision of two clumps and solve for the minimum core radius at which the
clump collision time is longer than the swarm dynamical time. Assuming
$N=M_{\mbh}/M_J$ clumps with collisional cross section
$\sigma_{\rm{coll}} = \pi (\Lambda_J/2)^2$, moving at speed $v_{\rm{swarm}} =
(R_c/2) \Omega_{\rm{swarm}}$ in a volume $\mathcal{V} = (4\pi/3) (R_c/2)^3$,
where $M_{\mbh}$ is the mass of the single BH formed by the swarm, we estimate
the collision time as,
\begin{equation}
t_{\rm{coll}} \equiv \frac{\mathcal{V}}{N \sigma_{\rm{coll}} v_{\rm{swarm}}} = \frac{M_J}{M_{\mbh}} \frac{4 R^2_c}{\pi \Lambda^2_J} \Omega^{-1}_{\rm{swarm}}
\end{equation}
(neglecting a factor of $3/\pi$).
Then the limit on the core radius, for which the
swarm dynamical time is shorter than the collision time, is
\begin{equation}
R_c \gtrsim \frac{1}{2}\sqrt{\pi  \frac{M_{\mbh}}{M_J}  } \Lambda_J \approx 5.2 \times 10^8 \rm{cm}  \ \left( \frac{M_{\mbh}}{30 \Msun }  \right)^{1/2} ,
\label{Eq:RcoreClump}
\end{equation} 
where we use the values for the Jeans mass and length above. This required
core size is consistent with the stellar size and the angular momentum budget
of Eq. (\ref{Eq:jojmx}).

It is also useful to compare the clump collision time with the free-fall time
in each Jeans unstable clump. The free-fall time of a clump is simply the
dynamical time in the clump, which we can relate to the dynamical time of the
entire swarm of clumps by $\Omega^{-1}_{\rm{clump}} = \sqrt{M_{\mbh}/M_J (\Lambda_J/R_c)^3}
\Omega^{-1}_{\rm{swarm}}$. Then the condition on the initial core radius that ensures that individual clumps collide before collapsing is,
\begin{equation}
R_c \lesssim \left(\frac{\pi}{4}\right)^{2/7}  \left(\frac{M_{\mbh}}{M_J} \right)^{3/7} \Lambda_J \approx 4.3 \times 10^8 \rm{cm}  \ \left( \frac{M_{\mbh}}{30 \Msun }  \right)^{3/7}.
\label{Eq:RclumpCollapse}
\end{equation} 
Putting together Eqs. (\ref{Eq:RcoreClump}) and (\ref{Eq:RclumpCollapse}), we
see that if the core is small enough for collisions to merge the clumps before
they are dynamically stirred, then collisions between clumps will also
occur before the clumps collapse. 

Because of the angular momentum budget of Eq. (\ref{Eq:jojmx}),
however, we expect that the core will not collapse to a size as small as the
limit in Eq. (\ref{Eq:RclumpCollapse}). More specifically, For stars with
radii greater than $\sim 4 \times 10^{10}$ cm, the core would acquire more
than the maximum break-up angular momentum when collapsing below the limiting
size in Eq. (\ref{Eq:RclumpCollapse}). Put another way, this limiting size is
approaching $\sim 50 r_G$ for a $60\Msun$ binary, at which point the core
radius is not much larger than the required binary separation. Hence, we favor
larger core radii, and thus the scenario where we are left with a swarm of
$\sim \Msun$ sized clumps that eventually merge due to gas plus
gravitational-radiation drag and collisions.

The timescale for the clumps of each swarm to merge into a $\sim30 \Msun$
BH, vs the timescale for the two ends of the dumbbell configuration to come
together is uncertain. To estimate an upper limit on the timescale for clumps
to form into the final 30 $\Msun$ BH, we compute the collision time of a swarm
of $\sim30$ Jeans-mass BHs with collisional cross section equal to the ISCO,
and in a region the size of half the core radius. This time is shorter than
the GW-decay time for a $30\Msun+30\Msun$ binary separated by the core radius,
regardless of the initial core radius.

Because the timescale for the swarm to be brought together via gas drag must
be at least a dynamical time for the stellar densities considered here and
because the clumps will collapse before colliding, we conclude that in this
fragmentation scenario the swarm of clumps will be able to stir itself
sufficiently to randomize the clump orbital angular momentum vectors
away from their birth directions before collisions and gas plus gravitational-
radiation drag collapse the swarm into a BH.

Then formation of the final BH from a part of this swarm could lead to
misalignment of the BH spin with the collapsing star's spin angular momentum,
and hence the eventual binary orbital angular momentum. The clumps that do not
form the final BH could escape (not greatly affecting the mass budget as the
stellar core can be much more massive than $60 \Msun$ \citep{Woosley:2016}) or
remain to impact either of the BHs in a collision that could misalign the BH
further, analogously to the processes that misalign the planets' spin axes
in the early solar system.

The evolution of the BH angular momentum due to clump collisions will follow a
random walk. The final angular momentum of the BH can be estimated from the
root mean square (rms) angular momentum delivered during the bombardment of
clumps with a given mass and velocity distribution. We use Eq. (20) of
\citep{Lissauer+1991} to make a pureley Newtonian estimate of the
expected rms angular momentum delivered to the BH, assuming only one impact
and assuming no angular momentum loss to gravitational radiation. To be
conservative we assume a clump mass $M_{\rm{clump}}$ equal to the Jeans mass
of Eq. (\ref{Eq:Jean}) (though the clump may have increased in mass between
collapse and impact) and a radius, $r_{\rm{clump}}$, equal to the clump
Schwarzschild radius (though the clump may be larger and hence deliver more
angular momentum). Then the rms angular momentum delivered by one impact and
written in terms of the total BH spin angular momentum before merger is,
\begin{widetext}
\begin{equation}
\frac{\Delta L}{L^i_{\mbh}} \approx 0.08 (\spin^{i})^{-1} \left( \frac{1 + \chi}{5/4} \right)^{1/2}   \left( \frac{ M_{\rm{clump}} }{ \Msun } \right)  \left( \frac{ M^i_{\mbh} }{ 29 \Msun } \right)^{-1} \left( \frac{ M^f_{\mbh} }{ 30 \Msun } \right)^{-1/2} \left( \frac{ 2r^i_{G} + r_{\rm{clump}} }{ r^f } \right)^{1/2} ,
\end{equation}
\end{widetext}
where $-1 \leq \spin \leq 1$ is the dimensionless BH spin parameter,
$L^i_{\mbh} \equiv \spin G (M^i_{\mbh})^2/c$ is the BH angular momentum,
$\chi$ is the squared ratio of impact speed to escape speed from the BH
(the speed of light), and the superscript $i$ ($f$) denotes the
quantity before (after) impact. Hence, the clump impacts can
alter the BH spin by $\sim 8 \%$ for an initially maximumly spinning BH. For a
two times more massive clump, and an initial BH spin of $\spin = 0.16$, the
above ratio reaches unity, and the clump impact could completely rearrange the
BH spin. Note that the above result implies that the BH spin would have a
value $\spin \sim N^{-1/2} \sim 0.2$ for $N\sim 25$, this is in agreement with
the observed spins of GW150914.

While the above processes could result in a BBH with misaligned spins,
they do not require it, they simply offer a channel for misalignment to
occur. Such a misalignment could lead to spin-orbital precession of the
binary. Precession could leave an observational imprint in the GW
\citep[\textit{e.g.},][]{Blanchet:2014LRR} and EM \citep{StoneLoebBerg:2013}
signatures of inspiral. However, precession could be problematic for jet
breakout \citep{ZhangWoosley+2004}, quenching the EM counterpart, or
shortening what would otherwise be longer bursts. There is presently
no strong evidence for precession in the LIGO data
\citep{GW150914:spinprec:2016, GW170104:2017}.

The final state of the collapsing clumps that make up the two  proto-BH swarms
is not clear. Given their initial compact size ($\Lambda_J$), they could
collapse to BHs, but future work needs to clarify this. If the clumps do not
collapse to BHs, then feeding of either proto-BH by clump collisions could
lead to large EM bursts that could clear out gas, possibly via jets,
from the core before merger of the final $30\Msun+30\Msun$ BBH.

If multiple clumps can collapse to black holes before collapse into one of the
components of the larger BBH, then mergers of smaller BHs within each end of
the rotating bar instability could generate non-standard GW signals in the
LIGO band \citep[See] []{Meiron+2017} prior to the main merger of the two
$\sim30 \Msun$ BBHs. While the timing of these non-standard GW signals
relative to the final merger is uncertain, we estimate that they would occur
before the main GW event by at least a merger time of the final 30+30$\Msun$
BBH system. Considering orbital decay due only to GWs and a $50 r_G$ initial
separation of the 30+30$\Msun$ system, we find that the putative smaller BBH
mergers would occur on order minutes before the main BBH LIGO signal. Mergers
of $1\Msun+1\Msun$ BHs, which have a $\sim30$ times smaller chirp mass, will
have a strain that is $\sim$two orders of magnitude lower than that of the
final $30\Msun+30\Msun$ merger. Mergers of $1\Msun+29\Msun$ or
$15\Msun+15\Msun$ BHs, however, would have $\sim25$ or $\sim3$ times smaller
strains respectively. Further study of this possibility may warrant a search
in the existing LIGO data.

Gravitational wave recoil kicks from these pre-mergers will depend on
the eccentricity and spins of the merging BHs, but for the final mergers of
$\sim1+30\Msun$ BHs, the sensitive mass ratio dependence of the kick velocity
makes the kick small. Using a maximum kick velocity of $\sim 4000$ km/s for
optimal spin alignment and mass ratio of $q\leq1=M_2/M_1 = 2/3$
\citep{Campanelli+2007_MaxRec}, the kick for a $q=1/30$ BBH drops by a factor
of $ q^2/(1+q)^5/0.035$ to $\lesssim100$ km/s. This is small compared to the
sound speed ($\sim 10^3$ km/s for $T\sim 10^9$ K) and also compared to the
binary orbital speed, even at a binary separation of $100M$ ($\sim 10^4$
km/s). We note that it is possible that the rare, largest possible kicks
between equal mass BHs in the swarm could marginally unbind them from the
swarm.

If such smaller BHs can form, the rate of mergers between smaller BHs or
between smaller BHs and the larger proto-BH in the LIGO band will depend upon
the redshift distribution of the single progenitor systems discussed here.

Finally, we note that the conditions in the collapsing star that lead
to fragmentation vs. direct collapse within each end of the rotating-bar
instability should be studied in future simulations which can capture the
effects of self-gravity and cooling needed to understand this process further
(e.g., in analogy to understanding a similar process in the context of
supermassive black hole seeds \citep{ChoiShlosBegel:2013}).

\subsection{Electromagnetic Emission}

We now consider the energetics and timescale of an EM counterpart of the BBH
merger. We carry out a calculation similar to that of Ref. \cite{Perna+2016}, but
in the setting of the single progenitor model.

Once the BHs form they will be driven together by gas torques, accretion, and
gravitational radiation losses \citep[\textit{e.g.},][]{FedrowOtt+2017}.
Accretion flows will form around each binary component and will be driven onto
the BHs via the magneto-rotational instability
\citep[MRI;][]{BalbusHawley:1991} and also spiral shock driven angular momentum
transport from disk perturbations due to the companion
\citep[see][]{Spruit+1987, Ryan+2017, JuStone+2017}. The outer edge of the
disk around each BH is given by the tidal truncation radius
\citep{Roedig+Krolik+Miller2014, Pac:1977},
\begin{eqnarray}
r^s_{\rm{out}} &\sim& 0.27 q^{0.3} a \\
r^p_{\rm{out}} &=& q^{-0.6} r^s_{\rm{out}},
\end{eqnarray}
which coincides with the location where orbit crossings exclude the
possibility of stable orbits at larger radii. Here $q=M_s/M_p$; $M_p>M_s$ is
the binary mass ratio, and $s$ and $p$ represent secondary and primary,
respectively.

The time for the material to be transported inwards to the BH from radius $r$ is
given by the viscous time there, 
\begin{eqnarray}
t^s_{\rm{in}} \equiv \frac{2}{3}\frac{r^2}{\nu} &=& \frac{2}{3}\frac{\HoR^{-2}}{\alpha} \sqrt{\frac{r^{3}}{G M}} \sqrt{1+1/q} \nonumber  \\
t^p_{\rm{in}} &=& q^{1/2}t^s_{\rm{in}},
\end{eqnarray}
where $M$ is the total binary mass and $\HoR$ is the dimensionless aspect ratio
(height over radius) of the disk. Here the $2/3$ prefactor is valid
for a steady-state disk and we have calculated the coefficient of kinematic
viscosity, $\nu$, using the Shakura-Sunyaev $\alpha$-prescription \citep{SS73}.

When the GW-decay timescale of the binary is longer than the viscous time at
the outer edge of the disk, the disk will evolve adiabatically and accrete at
the viscous rate onto each BH. However, when $t_{\GW} \leq t_{\rm{in}}$, at a
binary separation of
\begin{eqnarray}
\frac{a^s_{\rm{burst}}}{r_G} &\leq& \left(\frac{512}{15}\right)^{2/5} \frac{\HoR^{-4/5}}{\alpha^{2/5}} \frac{\left(0.27 q^{0.3} \right)^{3/5}}{\left(1+q\right)^{2/5} \left(1+1/q\right)^{1/5}}    \nonumber \\
a^p_{\rm{burst}} &=& a^s_{ \rm{burst} } q^{-0.16}  ,
\end{eqnarray}
the binary torque will drive the disk into the BH faster than the disk can
viscously respond and trigger a super-Eddington accretion event.
\footnote{See, for example, an analogue in the case of supermassive
black hole binaries: \citep[][and references
therein]{CerioliLodato:Squeeze:2016, Fontecilla+2017}.}

The resulting super-Eddington accretion burst occurs at time $t_{\rm{burst}} =
t_{\GW}(a_{\rm{burst}}$) before merger,
\begin{eqnarray}
t_{\rm{burst}} = 7.7 s \ \left( \frac{a_{\rm{burst}}}{24 r_G} \right)^4 \left( \frac{M}{60 \Msun} \right)^{-3} \left( \frac{(1 + q) (1 + \frac{1}{q})}{4} \right) , \nonumber \\
\label{Eq:tburst}
\end{eqnarray}
where $a=24 r_G$ corresponds to $a_{\rm{burst}}$ with $M=60 \Msun$,  $q=1$,
and fiducial, pre-burst disk parameters of $\HoR = 0.05$ and $\alpha =0.24$.

Given an efficiency $\eta$ for converting matter into energy, the luminosity
of the event is,
\begin{eqnarray}
\mathcal{L} &=& \eta \dot{M} c^2 \gtrsim \eta \frac{ r^3_{\rm{out}}(a_{\rm{burst}}) \HoR }{ t_{\rm{burst}} } \rho c^2 \nonumber \\
 &\gtrsim& 1.1 \times 10^{49} \rm{erg} \ \rm{s}^{-1} \left( \frac{\eta}{0.1} \right) \left(\frac{\rho}{ 10^8 \rm{g} \ \rm{cm}^{-3}} \right), 
\end{eqnarray}
where we use numbers corresponding to accretion onto the secondary BH, we
continue to use the fiducial disk parameters stated above, and the inequality
is written because the time of the accretion event must be less than
$t_{\rm{burst}}$ and we have not taken into account any beaming factors. Note
that for stellar core densities of $\rho \sim 10^{10} \rm{g} \ \rm{cm}^{-3}$,
even efficiencies of order $10^{-3}$ could still generate the observed
luminosities.

This luminosity is approximately $3 \times 10^9$ the Eddington value and will
drive a powerful outflow or relativistic jet. At the burst time of
approximately $8$ seconds before merger, given in Eq. (\ref{Eq:tburst}), this
outflow will clear out the gas surrounding the binary within a
sound crossing time,
\begin{eqnarray}
t_{\rm{clear}} &\lesssim& \frac{ a_{\rm{burst}} }{\HoR  v_{\orb}}  \\ \nonumber 
&\approx& 0.7 s \ \left( \frac{a_{\rm{burst}} }{ 24 r_G} \right)  \left(\frac{c/\sqrt{24}  }{v_{\orb}}\right) \left( \frac{\HoR}{0.05} \right)^{-1}.
\label{Eq:tclear}
\end{eqnarray}
We take this as an upper limit because the ambient sound speed is likely
larger than what we have assumed in the thin accretion flows around each BH.
Then the remaining $\sim 7$ seconds to merger will be unaffected by gas
torques and will not \citep[as suggested in Refs.][]{Dai+2016,
FedrowOtt+2017} affect the LIGO waveform which begins at $\sim 0.2$ seconds
before merger.

The quantity $t_{\rm{clear}}$ also provides an estimate for the duration of
the burst; once the gas is cleared from the binary orbit, the accretion event
will stop being powered. This $\lesssim 1$ second timescale is in
agreement with the observed durations of the GW150914 ($\sim 1$ second) and
GW170104 ($\sim 3.2 \times 10^{-2}$ seconds) gamma-ray transients.

There will be a delay between the super-Eddington accretion event plus jet
launching and the time at which the jet breaks out of the supermassive star,
generating the high-energy transient. \textcite{Woosley:2016} argued that the
stellar radius calculated in model R150A of Ref. \citep{Woosley:2016}, plus
the jet speed inside of the star calculated in Ref. \citep{ZhangWoosley+2004},
implies a delay of $\sim 10^{11} \rm{cm} / c/3 \sim 10$ seconds after
$t_{\rm{burst}}$, which yields a time of $\sim 2$ seconds after merger, and
because the GWs take $\sim 3$ seconds to reach the edge of the star as well, a
delay time between EM and GW emission of order $1$ second.

We point out that, within our model, jets could be launched from both
BHs. Simulations of super-massive BBH systems show that the jets launched from
each BH can combine into a single, larger jet near to the binary
\citep{Gold:GRMHD_CBD:2014}. A similar situation would be realized in our
model. This could result in a larger jet opening angle and higher probability of
observing the event along the jet axis than in the single-BH collapsar model.

While the fiducial system parameters chosen here yield a remarkable match to
the timescale observed between the GW and gamma-ray emission in GW150914 and
GW170104, we note that this delay timescale is highly dependent on system
parameters. The delay time depends on the gravitational wave decay timescale,
the critical binary separation at which the accretion event occurs, and the
radius and density of the collapsing star. In turn, these properties depend on
the binary mass, mass ratio, and the hydrodynamical properties of the accretion
flow around each black hole (parameterized by $\alpha$ and $\HoR$).

As an illustration of this parameter dependence, let us assume that the
secondary launches the observed jet with pre-burst accretion disk aspect ratio
$\HoR=0.05$, stellar breakout radius $10^{11}$ cm, and a jet speed inside the
star of $c/3$, then the predicted time lag for GW150914 is the observed $0.46$
seconds after merger if $\alpha = 0.275$. The predicted time lag for GW170104
is the predicted $0.4$ seconds before merger if $\alpha = 0.205$. We note that
this is in agreement with the values of $\alpha \sim 0.1 - 0.3$ expected
during the outbursting state of accretion onto BHs in cataclysmic variable
systems and also consistent with the values measured in simulations which
resolve the MRI \citep[see][and references therein]{JuStone+2017}).
Alternatively, if we assume a breakout radius of $7 \times 10^{11}$ cm
\citep{Bromm+2001}, and fix the pre-burst viscosity parameter to
$\alpha=0.24$, we find that $\HoR \sim 35.0$  to match the EM time delay for
GW150914, and $\HoR \sim 39.2$ to match the value for GW170104. Hence,
our model reproduces the observed EM-GW time delay when using standard values
of $\alpha$, $\HoR$, and the breakout radius. However, results are quite
sensitive to these parameters; precision to the third decimal in $\alpha$ and
to the first decimal in $\HoR$ is required to fix the delay time to the
reported hundredth of a second level precision. Reassuringly however, similar
parameters are required for both systems.

\section{Discussion and Implications}

We briefly compare our single progenitor model with related work in the
literature and then discuss implications of the model that can be used to test
it.

\textcite{Dai+2016} point out that, in models where the BBH orbits within the
stellar core, the orbital energy of the BBH will be converted into heat in the
surrounding gas via dynamical friction and could unbound the star before the
GRB-like event occurs. In our single progenitor model, we require a more
massive star than in Ref. \citep{Dai+2016}, having a higher binding energy and
a lower central density \citep{Fryer+2001}, causing the unbinding by dynamical
friction early in the BBH inspiral to be more difficult. Indeed for a central
stellar density of $10^{8}$ g cm$^{-3}$, and a $\gamma=2.5$ power law fall off
in the density of the stellar core (Eq. (1) of Ref. \citep{Dai+2016}), Figure
3 of Ref. \citep{Dai+2016} shows that the energy injected into the gas via
dynamical friction is below the binding energy of the progenitor star (even
for a progenitor half as massive as that considered here), as long as the
initial separation of the BBH is $\lesssim 10^{10}$ cm. This is in agreement
with our bounds on the initial binary separation in Eqs (\ref{Eq:jojmx}) and
(\ref{Eq:RcoreClump}). A final word on the fate of the gas in the vicinity of
the BBH before merger, however, must rely on more detailed calculations that
include heating and cooling of the gas and eventually radiation.

A few other scenarios have been put forth to explain a gamma-ray counterpart
to a BBH merger. \textcite{Woosley:2016} and \textcite{Janiuk+2017} envision a
close binary consisting of a BH and high mass star in which the BH spirals
into the star causing it to collapse into a BH. These scenarios could result
in a similar outcome as the single progenitor model; they do not, however,
provide a natural explanation for the near-unity mass ratios observed in
GW150914 and GW170401 and may be more susceptible to the unbinding of the star
as discussed above.

As noted, the model proposed in Ref. \citep{Perna+2016} for generating the
super-Eddington accretion event, is similar to that presented here, except that in
Ref. \cite{Perna+2016}, the gas needed for accretion is derived from a fossil disk
which slowly builds up in density as the binary comes together. In the fossil
disk scenario, the EM emission is prompt, not requiring time to break out from
a surrounding medium. Hence Ref. \citep{Perna+2016} uses $\HoR = 1/3$ and
$\alpha=0.1$ in order to cause the super-Eddington event to occur much closer
to merger. While the model of Ref. \citep{Perna+2016} hinges on the long term
survival and then slow pile up of this fossil disk, which has been disputed by
Ref. \citep{Kimura+2017}, it may still be viable and we discuss here the
predictions of the single progenitor model that would differentiate it from
alternate scenarios such as the fossil disk scenario:

\begin{itemize}

\item   
The systems envisioned here will not exist at the $\sim 10^3 r_G$
orbital separations that would be needed to place them in the high frequency
end of the LISA \citep{LISA:2017} band. We predict that LISA will not be
sensitive to BBHs in our single progenitor model, and hence the LISA
observations would derive a different BBH merger rate than LIGO as they will
probe a different population of BBHs. The single progenitor model presented
here could be ruled out if LISA and LIGO can link together GW observations of
a GW150914-like event \citep[\textit{e.g.},][]{Sesana:2016, Seto:2016} for
which gamma rays are detected near merger.

\item  
A low-luminosity supernova corresponding to the clearing of the gas in the
progenitor star envelope after the jet breakout should follow the GW and EM
signals in our single progenitor model. Similarly, the post merger
remnant could host a radio-afterglow \citep{Yamazaki+2016}. Future work
should address the observability of these signatures.

\item  
If the hydrodynamic properties of the accretion flow onto each BH, as well as
the stellar parameters, could be determined with better accuracy, then the
binary mass and mass ratio, measured from GWs, would allow us to predict the
EM and GW time delay and test the single progenitor model.

\item 
When the relativistic outflow is launched, the binary period is approximately
$0.2$ seconds. If the transient discussed here lasts of order one second, as
suggested by Eq. (\ref{Eq:tclear}) and the Fermi-GRB observation associated
with GW150914, then when the jet breaks out, its intensity would be modulated
due to the relativistic Doppler boost
\citep[\textit{e.g.},][]{PG1302Nature:2015b}, starting at a period of a
fraction of a second but chirping up in frequency by a few percent over $\sim
5$ orbits due to the orbital decay. If the EM chirp is detectable \citep[see
also][]{Schnittman+2017, Haiman:2017, HayLoeb:2016}, then it would constrain
the astrophysical factors which generate the EM and GW time delay discussed
above.

\end{itemize}

Furthermore, we make the following falsifiable statements pertaining to our model:
\begin{itemize}

\item{ 
Firstly, as we have stated, our model is sensitive to parameters. Because our
model can explain the gamma-ray emission from both GW150914 and GW170104 with
a narrow, self-consistent range of parameters, this implies that this
mechanism may only operate within this narrow range and that future events
should also be explainable by this narrow range.  Furthermore, if a gamma-ray
event indeed occurred for two out of the three LIGO events for which our
mechanism applies, and three out of five LIGO events are of the near-equal
mass, high mass BBH variety, then such gamma-ray counterparts should be
common.}

\item{ Our model involves a jet that must have a wide enough opening
angle to be detected in two out of three events. Because GW observations can
constrain the source orientation, future observation could falsify our model
via constraints on the jet scale.}

\item{ If each BH in the final BBH is formed from swarms of smaller
BHs, then LIGO, or future GW instruments should see this signal for sufficiently
nearby events.}

\end{itemize}

\section{Conclusions}

While the association between sub-second duration gamma-ray transients and the
merger of $30 \Msun$ BBHs is far from being firmly established, the now two
$\sim 3 \sigma$ detections of such transients within $0.5$ seconds of a BBH
merger motivates us to further examine the previously unexpected possibility
that BBH mergers can generate bright EM counterparts.

We have expanded upon the model of \textcite{Loeb:2016} for such an EM
counterpart to develop a scenario where bright EM emission from the more
massive GW150914- and GW170104-like BBH mergers is generated through a single
progenitor model. In the single progenitor model, the core of a very massive
($\sim 300 \Msun$), rapidly rotating star fragments via a rotational bar
instability and eventually forms two $\sim 30 \Msun$ BHs. At approximately
$10$ seconds before merger the BHs are fed by a burst of super-Eddington
accretion from the surrounding stellar-density matter due to the rapidly
increasing tidal torques of their companions. The accretion event can generate
$\gtrsim 10^{49}$ erg s$^{-1}$ luminosities during a powerful outflow that
clears the binary orbit of gas and launches a jet that breaks out from the
massive star within a few seconds of the merger, resulting in an EM and GW
time lag of $\lesssim 1$ second for the model parameters assumed here.

Whether or not this scenario reflects reality will ultimately be tested with
future LIGO observations and their EM follow up, as well as multi-band GW
observations with the upcoming LISA mission. Future gamma-ray plus BBH merger
associations will warrant further, more detailed analysis of the model
presented here.

\acknowledgements
The authors thank Jeffrey J. Andrews, Konstantin Batygin, Edo Berger, and Nick
Stone for useful discussions and comments. The authors also thank the
anonymous referees for comments which improved the manuscript. Financial
support was provided from NASA through Einstein Postdoctoral Fellowship award
number PF6-170151 (DJD). This work was supported in part by the Black Hole
Initiative, which is funded by a grant from the John Templeton Foundation.

\bibliography{refs}
\end{document}